# CHEMICAL GAS SENSORS BASED ON NANOWIRES

**_Yaping Dan[1], Stephane Evoy[2] and A. T. Charlie Johnson[1,3]_**

[1]Department of Electrical and Systems Engineering, University of Pennsylvania, Philadelphia, PA 19104, USA

[2]Department of Electrical and Computer Engineering and National Institute for Nanotechnology, University of Alberta, Edmonton, AB T6G 2V4, Canada

[3]Department of Physics and Astronomy, University of Pennsylvania, Philadelphia, PA 19104, USA

## ABSTRACT

Chemical gas sensors based on nanowires can find a wide range of applications in clinical assaying, environmental emission control, explosive detection, agricultural storage and shipping, and workplace hazard monitoring. Sensors in the forms of nanowires are expected to have significantly enhanced performance due to high surface-volume ratio and quasi-one-dimensional confinement in nanowires. Indeed, chemical gas sensors based on nanowires with a ppb level sensitivity have been demonstrated. In this review, the fundamental aspects on (i) methods of nanowire synthesis (ii) performance of nanowire sensors, (iii) chemiresistors, transistor sensors, and their sensing mechanism, and (iv) assembly technologies will be summarized and discussed. The prospects of the future research on chemical gas sensors based on nanowires will be also addressed.

## 1. INTRODUCTION

Today's computer technologies have already dwarfed the capability of the human brain in many aspects. Current visual technologies have also surpassed some functions of the human eye. However, despite several decades of intensive research, the goal of creating an artificial "electronic nose" (e-nose) that can compete with a biological olfactory system has yet to be achieved. The difficulties to make sophisticated electronic noses lie in the two aspects. First is the extremely large chemical diversity and massive parallelism that is characteristic of mammalian olfactory systems. For example, the human nose has more than 400 different types of sensing cells and each type is replicated over 100,000 times, for a total of around forty million cells overall. The microelectronic technology faces a huge obstacle to make such a sensor array in terms of the scale and chemical diversity. A second important factor is that microscale chemical sensors are typically not as sensitive as their counterparts in biological olfactory systems. For example, human noses can detect analytes as low as tens of ppb [1] that microsensors usually cannot.

Recent developments in nanotechnology offer the possibility of overcoming these challenges and creating a path to functional e-nose systems whose performance rivals that of biological olfaction. Nanoscale chemical sensors based on one-dimensional nanostructures (nanotubes, nanowires, nanofibers, etc.) have been demonstrated to be excellent candidates for use as chemical sensors because of the enhanced sensitivity that derives from their very high surface-to-volume ratios. For example, $In_2O_3$[2,3], Si[4] and $V_2O_5$[5] nanowires with a diameter smaller than 25nm are able to sense 5ppb $NO_2$, 20ppb $NO_2$ and 30ppb 1-butylamine, respectively. Such structures have been made from a broad array of materials (semiconductors, oxides, polymers, and metals), implying that broad chemical diversity might be achieved. In addition, significant technical progress has been made in the use of non-traditional fabrication approaches (e.g., patterning by diblock copolymer self-assembly [6,7], nano-imprint lithography [8], and dielectrophoretic assembly [9]) that may enable the creation of nanosensor arrays with unprecedentedly high density.

The goal of this paper is to provide an overview of the rapid progress in this research area. By organizing, comparing and evaluating existing approaches, we aim to define a clear vision of the future research direction.

In this review, we will survey a portion of this broad research field, and to evaluate progress in the various areas with a focus on: (i) methods of nanowire synthesis (ii) performance of nanowire sensors, (iii) chemiresistors, transistor sensors, and their sensing mechanism, and (iv) assembly technologies. The prospects of the future research on chemical gas sensors based on nanowires will also be addressed.

## 2. NANOWIRE SYNTHESIS METHODS

More than ten distinct methods of nanowires synthesis have been demonstrated to date [2,10-22]. Here, we focus on two common and versatile methods that are widely used to synthesize nanowires for sensing applications: the vapor-liquid-solid method (VLS) and the templating method. The vapor-liquid-solid method is typically accomplished in a low pressure, high temperature furnace. It has been employed to synthesize metal oxide [2,3,11,23-31] and semiconducting nanowires [4,14,32]. The templating method is based on electroplating of materials into a template structure consisting of aligned parallel nanopores. ”Indirect” methods of electroplating have been developed for materials (e.g., metal oxides) whose low conductivity precludes direct use of electroplating, as will be discussed below. In addition, “Directed Electrochemical Nanowire Assembly”, an interesting nanowire synthesis technique that was recently discovered, will be introduced at the end of this section.

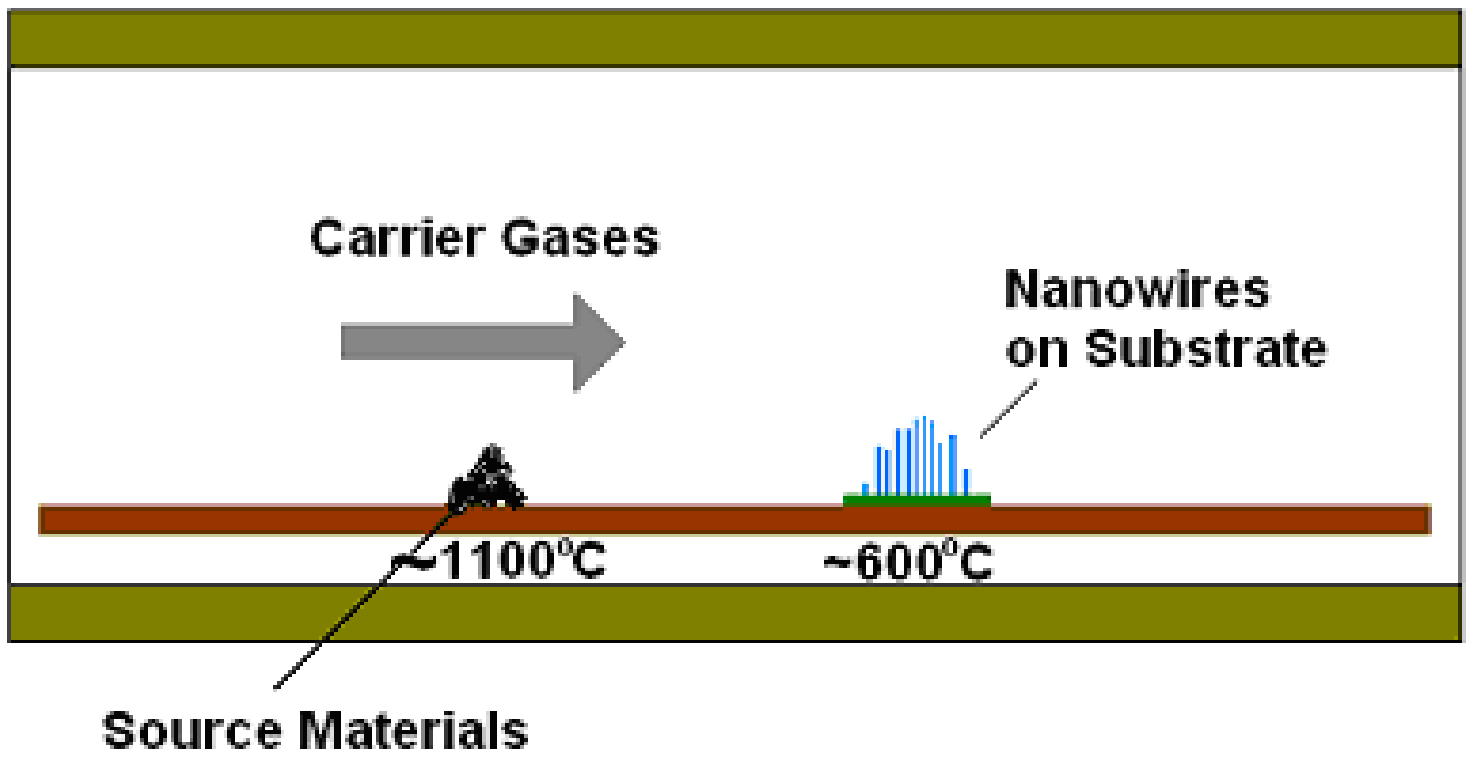

Figure 1. The furnace diagram of LVS nanowire synthesis method.

## 2.1 Vapor-Liquid-Solid Method

Vapor-liquid-solid (VLS) method is a commonly used procedure to synthesize nanowires. Nanowires consisting of $In_2O_3$[2,3], $Ga_2O_3$[9,33], $SnO_2$[10,23-25], ZnO[26-31], $WO_3$[34], $TeO_2$[35], $V_2O_5$[13,5], $ZnSnO_3$[36,37], Ge[38] and Si [4,14,32] have been grown using this method, among the other materials. The growth process is typically accomplished in a low pressure, high temperature furnace (Figure 1). The temperature near the source is elevated sufficiently to melt the source material so they may evaporate. A carrier gas flow brings the vapor to the substrate where nanowires grow with the assistance of catalysts. The catalyst material may be pre-deposited on the growth substrate, or it may form spontaneously during the VLS growth process, as described below.

VLS growth methods can be categorized according to the dominant physical-chemical growth process and the growth system. In terms of process, it can be classified as metal catalyst or non-metal catalyst (e.g., oxide or sulfide) VLS. The growth system used is typically either thermal evaporation, laser ablation or inductive heating assisted synthesis.

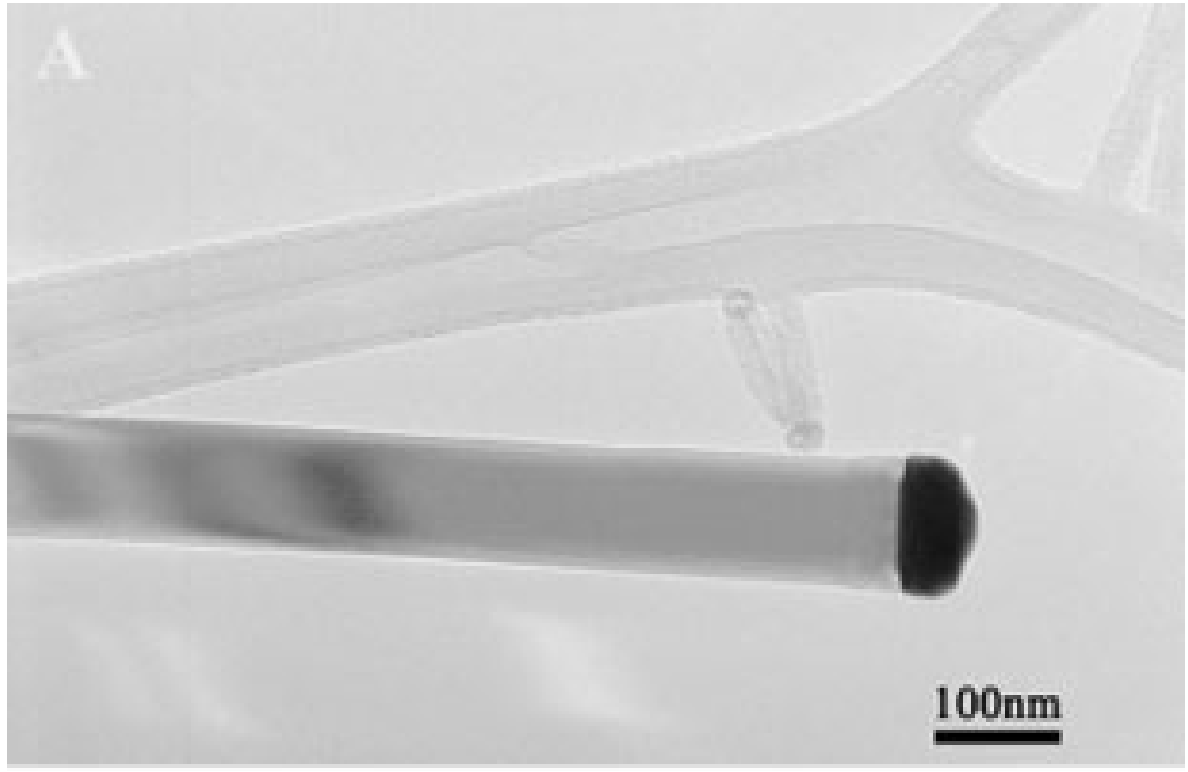

Figure 2. A gold catalyst particle on the tip of a ZnO nanowire (from Ref. [39]).

The metal catalyst VLS method uses metals such as Au, Fe, Co and Ni as catalysts. The metal catalysts can be mixed with the source material or spread on the substrate where the nanowires grow. In either case, the metal catalyst is either patterned or self-organized into nanoparticles (NPs). These NPs react with the source vapor forming solution droplets on the substrate serving as a preferential site for absorption of reactant, since there is a much higher sticking probability on liquid vs solid surfaces. When the droplets become supersaturated, they are the nucleation sites for crystallization. Preferential 1D growth occurs in the presence of reactant as long as the catalyst remains liquid. During this process, the catalyst particle tends to remain at the tip of the growing nanowire (Figure 2). See Ref. [39,40] for details.

The size of the catalyst particles that are used to generate the nanowires depends on the preparation process. Typical methods include thin film deposition of the metal catalyst on the substrate by thermal evaporation or sputtering [25,41]. The metal thin film will cluster into small particles when heated up to the growth temperature. This typically leads to a wide distribution in NP diameter that is reflected in the diameter distribution of the resultant nanowires. The second approach is to deposit prefabricated monodisperse catalyst nano-particles on the substrate. Since the prefabricated nano-particles are uniform in size, nanowires can grow more uniformly in diameter [3].

Commonly used carrier gases include argon and nitrogen. Oxidizing gases may be mixed in the carrier gas, depending on the source material and the desired composition of the nanowires. For example, when growing metal oxide nanowires with the metal powder source, $O_2$ is often mixed in the carrier gas [26].

Oxide-assisted and sulfide-assisted growth are non-metal catalyst VLS which have been reported to prepare Si [42], GaAs [43], MgO [44] nanowires. The oxide (or sulfide) played a critical role through the nanowire growing process. For example, it was observed that source material consisting of silicon blended with silicon oxide led to the growth of high-quality silicon nanowires [42], but growth of the resulting nanowires could not be continued using a pure silicon source. Although the exact mechanism of this synthesis remains unknown, the following explanation is believed to be plausible [42]. The sub-oxide $SiO_x$ ($x<2$) vapor from the silicon source blended with $SiO_2$ is liquefied and eventually becomes supersaturated at the cooler substrate. Oxygen atoms are expelled from the supersaturated liquid and form a silicon dioxide shell layer surrounding a pure silicon crystal. However, the tip of the nanowire, somehow, remains silicon sub-oxide $SiO_x$ ($x<2$), which leads to the continuing (directional) growth of nanowires, while $SiO_2$ on the shell stops the nanowire from growing laterally. A pure silicon source without the $SiO_2$ catalyst cannot continue this process.

If we categorize the LVS method in terms of the preparation system, it can be classified as thermal evaporation [10,24,25], laser ablation [2] and inductive heating assisted synthesis [45], which differ only in the heating source. Some of the heating sources have advantages over the others. For example, laser ablation may enhance the uniformity of the nanowire diameter [2]. The inductive heating assisted synthesis can be more effectively heating up samples than the conventional thermal evaporation, which leads to a much shorter synthesis time [45].

## 2.2 Templating Method

The templating method is a second powerful way to synthesize nanowires for use in electronic devices. The idea is to fill materials into a nanoporous template (mostly by electrochemistry) to form nanostructures that can be released by dissolving the template. Nanoporous anodic aluminum oxide (AAO) membranes are the most extensively used templates for nanowire synthesis [11,46,47, 13]. The AAO membranes are fabricated by anodizing aluminum foils in acids at a voltage higher than 15V [59]. The pores inside the membrane are perpendicular to the surface and positioned in the form of a hexagonal lattice (See figure 3a). The pore size is proportional to the anodizing voltage. Currently the AAO membranes are commercially available from Whatman, Inc.

To synthesize nanowires, materials have to be filled into the nanopores in some way. Electrochemistry is a powerful method for such applications and has been used to synthesize nanowires consisting of metals [60-66], conducting polymers [9], semiconductors [49] and nonconductive metal oxides [11,13,46-48]. It is straightforward to fill metals and conducting polymers into the template by electrochemistry, while semiconducting and non-conducting materials can only be filled into the nanopores in an indirect way, as described in the following paragraph.

One "indirect" method for electrochemical deposition of non-metals is the so-called sol-gel technique, which has been used to synthesize nanowires from titanium oxide [48] and CdS [50], among other materials. Here we take the synthesis of titanium oxide nanowires as an illustrative example. A sol-gel is a colloidal suspension. It has been shown that the solution $TiO(SO_4)$ and a sol-gel form of $TiO(OH)_2{\cdot}x(H_2O)$ can be switched back and forth by increasing (adding $NH_3{\cdot}H_2O$) or decreasing (adding $H_2SO_4$) the pH value above or below 3, shown in Figure 4. We first prepare a solution of $TiO(SO_4)$ with a pH value of approximately 2.5 [48]. When an appropriate voltage is applied cross an AAO membrane (Figure 3b) in the solution, the local pH value inside the nanopores near the cathode will increase above 3, which turns the solution into a sol-gel form of $TiO(OH)_2{\cdot}x(H_2O)$. This process continues till the pores are fully filled with the sol-gel. After annealing at temperatures greater than 450ºC, the sol-gel turns into $TiO_2$ crystal nanowires that can be released later by dissolving the membrane in KOH. In addition to one dimensional nanowires, we have synthesized more complicated 3-D periodical structures for photonic applications using 3-D templates based on SU-8[51,52].

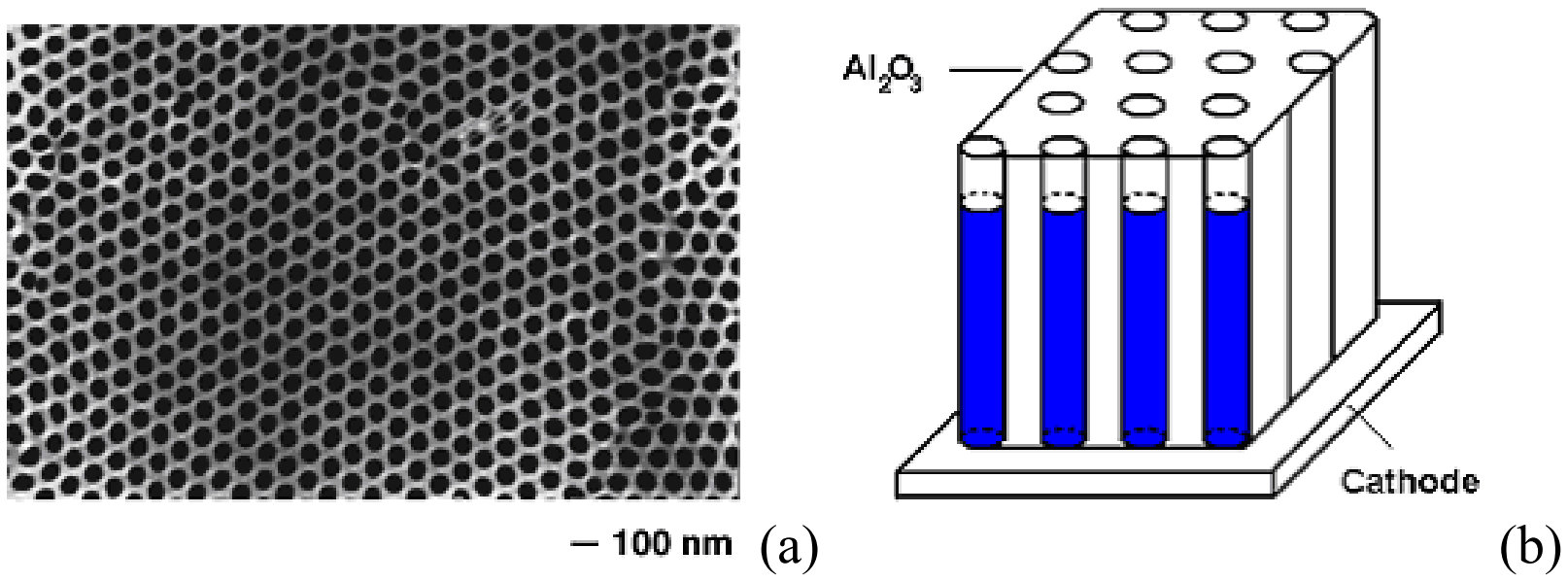

Figure 3. (a) The SEM image of AAO membrane surface (from Ref. [59]) and (b) AAO membrane used a template to make nanowires by electroplating.

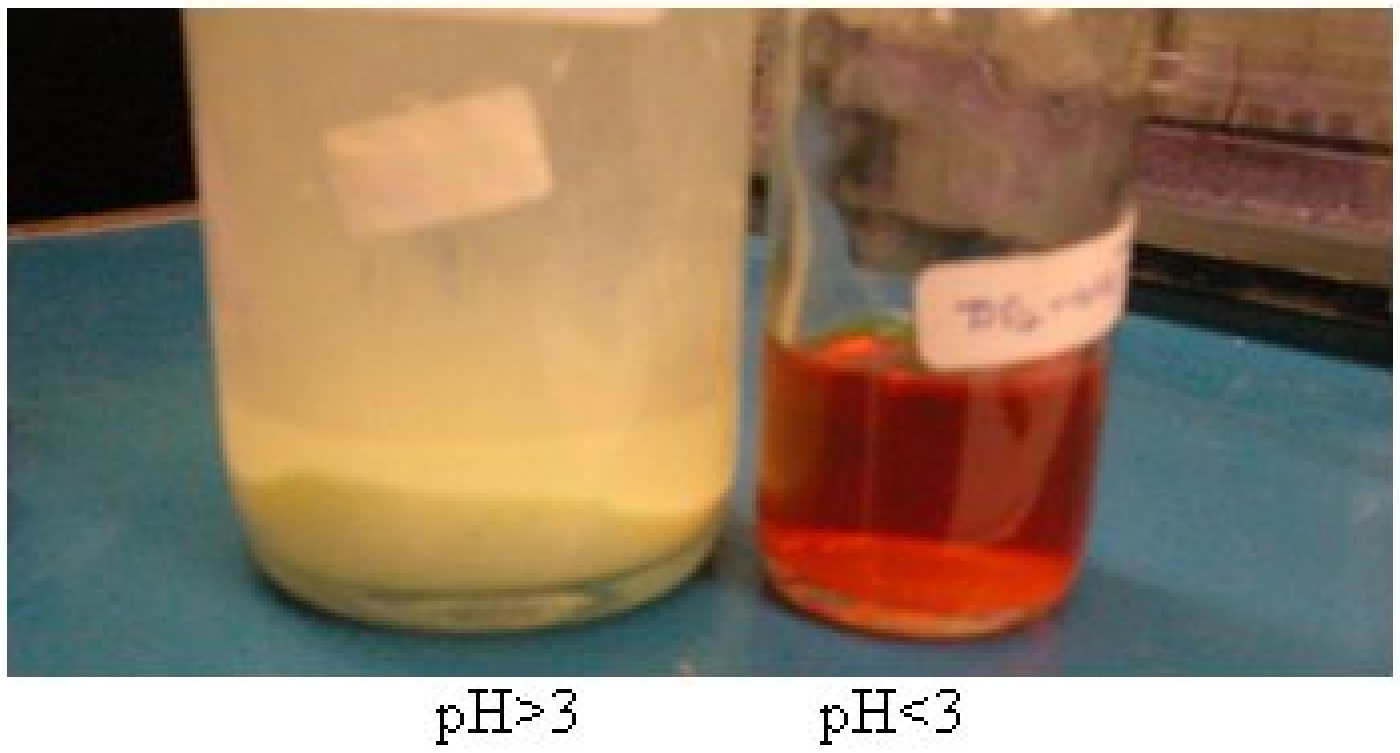

Figure 4. $TiO(OH)_2 \cdot x(H_2O)$ sol-gel and the $TiO(SO_4)$ solution.

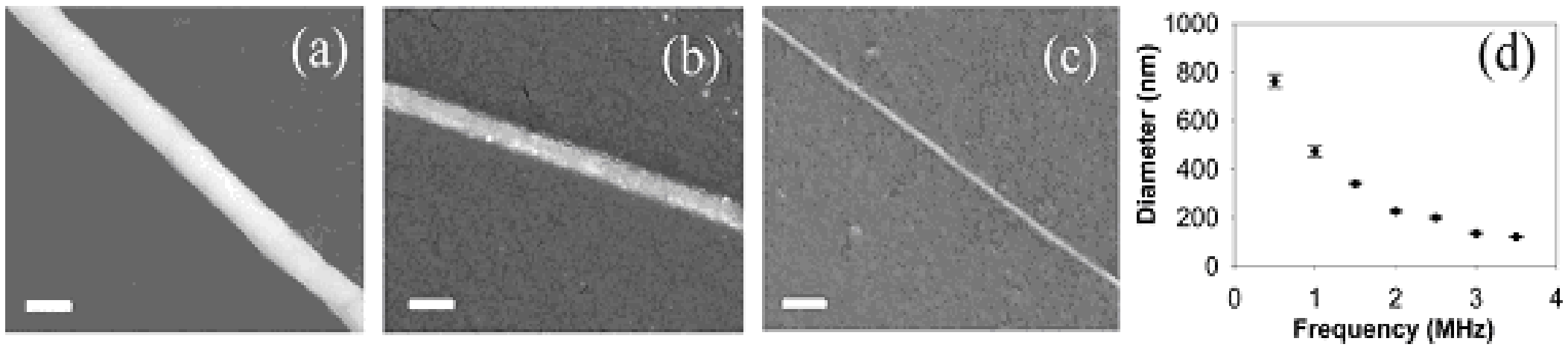

Figure 5. SEM micrographs of wires grown from 0.055 M $In(CH_3COO)_3$ solutions with voltage frequencies of (a) 0.5 MHz, (b) 1.0 MHz and (c) 3.5 MHz. The scale bars represent 1 μm. (d) Wire diameter as a function of the frequency of the alternating voltage (Ref. [54]).

$$TiO(OH)_2 \cdot xH_2O(sol) \underset{NH_4OH}{\overset{H_2SO_4}{\Longleftrightarrow}} TiO(SO_4)$$

There are also some other techniques to prepare metal oxides by electro-chemistry. For example, $Zn(OH)_x$ can be formed through electroplating $Zn(NO_3)_2$ solution at an elevated temperature in the presence of $O_2$ [53]. The $Zn(OH)_x$ can also be filled into nanoporous templates and form ZnO crystal nanowires after a high temperature annealing.

## 2.3 Directed Electrochemical Nanowire Assembly

Directed electrochemical nanowire assembly (DENA) [54] is a recent, very interesting approach to nanowires growth that involves voltage-induced crystallization of metallic wires from aqueous salt solutions. The mechanism behind this is dendritic solidification. A characteristic of dendritic solidification is that the growth velocity and tip radius are anti-correlated, which is exploited to realize diameter-tunable nanowire growth. The experimental parameter that provides this control is ω, the frequency of the alternating voltage. Increasing ω effectively steepens the metal cation concentration gradient at the wire-solution interface, thereby increasing the growth velocity of the wire. Figure 5 shows In nanowires induced by a square-waveform signal at different frequencies. As the frequency increases from 0.5 to 3.5

MHz, In wires exhibit a diameter ranging from 150nm down to 45nm. This method can easily control the diameter of the nanowires and grow nanowire arrays without any template.

In addition to these three methods, other methods were also used to synthesize nanowires such as aqueous-solution growth [20], molecular beam epitaxy [12,55], electrospinning [5] and seed-mediated growth [19].

## 3. PERFORMANCE OF NANOWIRE SENSORS

Surface states play a very important role in the response of solid-state chemical gas sensors [10,24,25,27,29,56,57]. Sensors in nano scale significantly enhance surface/volume ratio, which augments this role and brings many benefits to the three “S” (sensitivity, selectivity and stability) of sensor technology [58]. Nanowires are an ideal candidate for this application due to their additional advantages such as simple methods for synthesis and contact fabrication, and one-dimensional confinement of carriers. In recent years, nanowire sensors made from materials such as metal oxides [2-37], metals [60-66], silicon [4,32] and conducting polymers [9,67] have been intensively investigated. We provide a brief review of the progress here, with a summary found in Table I at the end of this chapter.

### 3.1 Metal Oxide Nanowire Sensors

Sensitive metal oxide sensors typically have to operate at an elevated temperature, which results in large power consumption and a complicated measurement system, especially for metal oxide nanowires that are much more resistive.

However, metal oxides have long been used as sensors due to their high sensitivity and stability [2,3], and such sensors have been commercially available for more than 30 years. It is interesting to take advantage of the merits of the nano-size feature to improve their performance. This fact likely explains the large amount of research that has been done to explore the sensing properties of nanowires formed from such materials. The performance of nanowire sensors can be significantly improved only when the diameter is smaller than 25nm [68] that is comparable to the grain size in thin film sensors. In term of low detection limit, $In_2O_3$ nanowires with 10 nm diameter, synthesized using the VLS method, were reported to have the best performance [2-3]. Both single and multiple nanowire sensors were made, contacted by gold electrodes, and used in a field effect transistor (FET) configuration to sense analytes at room temperature (Figure 6a and b). The data showed the multiple nanowire sensors had a sensitivity 4-5 orders of magnitude better than that of thin film sensors of the same material. As an example, exposure to 5ppb $NO_2$ at room temperature induced a 20% change in conductance, although this response was rather slow, requiring 1000s (Figure 6c). The single nanowire device had an inferior performance. This was tentatively attributed to enhanced sensitivity associated with nanowire/nanowire junctions, found only in multiwire devices.

**Table 1. Summary of Research on Nanowire Gas Sensors**

| Material | Device | Diameter (D) | Analytes | Detect Limit | Resp Time | Work Temp | Synthesis | Sensitivity, Selectivity and Comments | Reference |
|---|---|---|---|---|---|---|---|---|---|
| $In_2O_3$ | FET | 10nm | $NO_2$ , $NH_3$ | $NO_2$ 0.5ppm | | Room Temp (RT) | Laser Ablation | 4-5 orders more sensitive than thin films | [2] Appl.Phys.Lett. 82(2003)p1613 C.Zhou |
| $In_2O_3$ | FET | 10nm | $NO_2$ $NH_3$ $O_2$ CO $H_2$ | $NO_2$ 5ppb | 5-10s | RT | Laser Ablation | Doping improves Selectivity for $NH_3$ | [2] Nano Lett. 4(2004)p1919 C.Zhou |
| $In_2O_3$ | FET | 10nm | $NH_3$ | 0.002% $NH_3$ | 20s | RT | CVD | | [3] J Phys Chem B 107(2003)p12451 C.Zhou |
| $Ga_2O_3$ | Resistor | Nanowire thin film | | | | 600C | Ga Oxidized | Ti-doped improves sensitivity to 6% | [9] J Mat. Research 19(2004)p1105 C.Chen |
| $Ga_2O_3$ | Resistor | 50-90nm | Ethanol | 1500ppm | | 100C | Ga oxidized in water vapor | | [33] IEEE Sensors, 5(2005)p20, M.F.Yu |
| $SnO_2$-In | Resistor | 70-150nm | Ethanol | 10ppm | 2s | 400C | Thermal Evaporate | | [10]Appl.Phys.Lett. 88(2006)p201907 T.H.Wang |
| $SnO_2$ | Resistor & Photolum | | $O_2$, CO, $NO_2$, Ethanol | CO 10ppm $NH_3$ 50ppm $NO_2$ 1ppm | 30s | 200C | Vapor Phase Deposition | PL highly selective to $NO_2$ | [23] Sens. & Actuat B 109(2005)p2 C.Baratto |
| $SnO_2$-Sb | Resistor | 40nm | Ethanol | 10ppm | 5s | 300C | Thermal Evaporate | Sb-$SnO_2$ : low resistance & quick response | [24] Chem Commun. 2005, p3841 Q. Wan |
| $SnO_2$-Ru | Resistor | 100-900nm | $NO_2$, Liquid Petroleum Gas | 50ppm | 30-90s | 250C | Thermal Evaporate | Ru-$SnO_2$ : highly selective to $NO_2$ at Room Temp (RT) | [25] Sens. & Actuat B 107(2005)p708 I.S. Mulla |
| $SnO_2$ | FET | 60nm | $O_2$, CO | | 200s | 550K | Electro-deposition | Surface quality has huge effects on sensitivity | [11] J Phys Chem B 109(2005)p1923 A.Kolmakov |
| $SnO_2$ | FET | 60nm | $O_2$, CO | | | 400-500K | Electro-deposition | | [46] Nano Lett. 3(2004)p403 A. Kolmakov |
| $SnO_2$ | Resistor | 60nm | $O_2$, CO | | 40s | 500K | Electro-deposition | | [47] Adv. Mater 15(2003)p997 M.Moskovits |
| $SnO_2$ | Resistor (AC) | 20nm | CO | CO 5ppm Stability 4% | 100s | 20-200C | | Diameter D<25nm needed for higher sensitivity | [68] Sens. & Actuat B 121(2007)p3 J.R. Morante |
| ZnO | FET | | $NH_3$, $NO_2$ | $NO_2$ :200ppb $NH_3$ : 0.5% | 100s | | | Backgate refreshable | [98] Appl. Phys. Lett. 86(2005)p123510 J.G. Lu |
| ZnO | FET | 40-60nm | $O_2$ | | | | Thermal Evaporate | | [26] Appl. Phys. Lett. 85(2004)p6389 T.H. Wang |
| ZnO-Pd, Pt, Au, Ni,Ag,Ti | Resistor | 30-150nm | $H_2$ | 10ppm | 10min | RT | Thermal Evaporate | Pt: 8% change in R; Pd: 2 times; others 10 times | [27] Appl. Phys. A 81(2005)p1117 S. J. Pearton |
| ZnO | Resistor | 30nm | Ethanol | 50ppm Ethanol (18% Change in R) | 60s | RT-300C | Thermal Evaporate | Most sensitive at 300C | [28] Sens. & Actuat B (2007) I-Cheng Chen |
| ZnO:Ga | Resistor | 50-125nm | CO | | | 320C | Thermal Evaporate | | [29] Sens. & Actuat B 125(2007)p498 I-Cheng Chen |

**Table 1. Continued**

| Material | Device | Diameter (D) | Analytes | Detect Limit | Resp Time | Work Temp | Synthesis | Sensitivity and Selectivity | Reference |
|---|---|---|---|---|---|---|---|---|---|
| ZnO | FET | 60nm | $O_2$ | 10ppm | | RT | Vapor-Liquid-Solid | Smaller diameter more sensitive | [30] Appl. Phys. Lett. 85(2004) p5923 J.G. Lu |
| ZnO | Quartz Crystal Micro-balance | 20nm | $NH_3$ | 40-1000ppm | 5s | RT | Thermal Evaporate | | [31] Appl. Surf. Sci. 252(2006) p2404 J. Zhang |
| ZnO-Pd | Resistor | 30-150nm | $H_2$ | 10ppm | 20s | RT | | Pt doped samples 5 times more sensitive | [56] Appl. Phys. Lett. 86(2005)p243503 S.J. Pearton |
| ZnO | Resistor | 30-150nm | $H_2$, $O_3$ | 3% $O_3$ | | 112C for $H_2$; RT for $O_3$ | MBE | | [12] Appl. Phys. A 80(2005)p1029 S.J. Pearton |
| ZnO-Pt | Resistor | 30-150nm | $H_2$ | 500ppm | 10min | RT | MBE | Pt can increase sensitivity 10 times | [55] Electrochem & solid state lett 9(2005)G230 S. J. Pearton |
| $WO_3$ | Resistor | 100nm | $NO_2$, $H_2S$ | $NO_2$: 50ppb $H_2S$: 10ppm | Several mins | 250C | Thermal Evaporate | Annealing Improves sensitivity 20 times | [34] Appl. Phys. Lett. 88(2006)p203101 Z.L. Wang |
| $TeO_2$ | Resistor | 30-200nm | $NO_2$, $NH_3$, $H_2S$ | $NO_2$ 10ppm $NH_3$ 100ppm $H_2S$ 50ppm | 10min | RT | Thermal Evaporate | | [35] Appl. Phys. Lett. 90(2007)p173119 Z. Liu |
| $V_2O_5$ | Resistor | | He | | | | Electro-phoresis Sol-gel | | [13] Appl. Phys. Lett. 86(2005)p253102 H.Y. Yu |
| $V_2O_5$ | Resistor | 10nm | 1-butylamine toluene 1-propanol | 1-butylamine: 30ppb | 500s | RT | Electro-spinning | Good for medical applications | [5] Sens. & Actuat B 106(2005) p730 T. Vosseyer |
| $ZnSnO_3$ | Resistor | 20-90nm | Ethanol | 1ppm | 1s | 300C | Thermal Evaporate | Sensitivity 42 at 500pm ethanol | [36] Appl. Phys. Lett. 86(2005)p233101 T.H.Wang |
| $ZnSnO_3$ | Resistor | 50nm | $O_2$ | Much more sensitive than ZnO, ZnS, $Ga_2O_3$ | | | Thermal Evaporate | | [37] Appl. Phys. Lett. 91(2007)022111 T.H. Wang |
| $LiMo_3Se_3$ | Resistor | 4-6nm | Methanol, THF, Acetonitrite DMSO | DMSO:2.6ppm MeCN:520ppm MeOH:130ppm | Several seconds | RT | | | [99] JACS Comm. 127(2005) p7666 F.E. Osterloh |
| $LiMo_3Se_3$-Lithium Iodide, CTA | Resistor | 5nm | | | 5s | | | | [100] Langmuir 22(2006) p8253 F. E. Osterloh |
| Ag | Resistor | 150-950nm | $NH_3$ | | 5s | RT | Electro-deposition | 10000% change in R for $NH_3$; no Response to CO, $O_2$, $CH_4$, Ar, water | [60] Nano Lett. 4(2004) p665 R.M. Penner |
| Au | Gas Ionization Sensor | 180nm | | | | | Electro-deposition | | [61] Sens. & Actuat B 137(2007) p248 R.B. Adeghian |
| Au-porous | Resistor | 100nm | ODT | | | RT | Electro-deposition | Sensitivity comparable to thin films | [62] J Phys. Chem. B 110(2006) p4318 P.C. Searson |

**Table 1. Continued**

| Material | Device | Diameter (D) | Analytes | Detect Limit | Resp Time | Work Temp | Synthesis | Sensitivity, Selectivity and Comments | Reference |
|---|---|---|---|---|---|---|---|---|---|
| Pd | Resistor | 75nm | $H_2$ | 0.02% | 370s | RT | Electro-deposition | | [63] Small 2(2006) p356 M. Yun |
| Pd | Resistor | Mesowire | $H_2$ | 0.5% | 20ms-5s | RT | Electro-deposition | Imprvd Perfmce with presnce of $O_2$, $H_2O$, $CH_4$ and CO | [64] Anal.Chem. 74(2002) p1546 R. M. Penner |
| Pd | Resistor | 250nm | $H_2$ | 2% | 1s | RT | Electro-deposition | | [65] Sens. & Actuat B 111(2005) p13 M. Z. Atashbar |
| Pd | Resistor | 200nm | $H_2$ | 0.5% $H_2$ | <75ms | RT | Electro-deposition | | [66] Science 293(2001) p2227 R. M. Penner |
| Si | FET | 18nm | $NO_2$ Acetone Hexane | 20ppb $NO_2$ | 1min | RT | Photo-lithography Etch | Comparable to Carbon NT, Ceramic sensors | [4] Nature Mater. 6(2007) p379 J. R. Heath |
| Si | Resistor | 20nm | $NH_3$ | 0.1% | Mins | | Oxide-Assisted Growth | | [14] Chem. Phys. Lett 269(2003) p220 S. T. Lee |
| Si | FET | 40nm | $O_2$ | 9% decrease in R after 20% $O_2$ exposure | Mins | RT | Confined Lateral Selective Epitaxy | | [32] Appl. Phys. Lett. 83(2003) p.4613 R. Bashir |
| PANI | Resistor | 30-50nm | HCl Vapor, $NH_3$ | HCl: 100ppm $NH_3$: 100ppm | | RT | Chemical oxidative Poly-merization | Sensitive than thin films | [67] JACS Comm. 125(2003) p314 R. B. Kaner |
| PANI | Resistor | 30-120nm | HCl, $NH_3$, $N_2H_4$, $ChCl_3$, $Ch_3OH$ | 3ppm | 2s | RT | Interfacial poly-merization | Sensitivity higher than thin films | [71] Nano Lett. 4(2004) p491 R. B. Kaner |
| PANI | Resistor | 40-80nm | HCl, $NH_3$ | $NH_3$ 0.5ppm HCl 100ppm | 5s | RT | Electro-deposition | 4 orders change in R in 5s | [73] Nano Lett. 4(2004) p1693 J. R. Heath |
| PANI | Resistor | 100nm | $NH_3$ | 0.5ppm | 10s | RT | Electro-deposition | Sensitivity depends on the diameter | [74] Nano Lett. 4(2004) p671 H. G. Craighead |
| Polymer | Resistor | 80-180nm | $NH_3$ | 1ppm | | RT | Electro-deposition | | [101] Chem. Comm. (2006) p3075 H. R. Tseng |
| Polymer | Resistor | 50-300nm | | | | RT | Electro-spinning | | [15] J Mater Chem. 14(2005) p1503 J. Kameoka |
| Polymer-wire | Tuning Fork Fre-quency | 100nm-500nm | Ethanol | 15ppm | 30s | RT | Stretch Pull | | [16] Nano Lett. 3(2003) p1173 N. J. Tao |
| PEDOT /PSS | Resistor | 200nm | Ethanol Methanol Acetone | <100ppm | 30s | RT | Electro-deposition | Respond faster than thin films | [9] Sens. & Actuat B 125(2007) p55 Y. Dan |

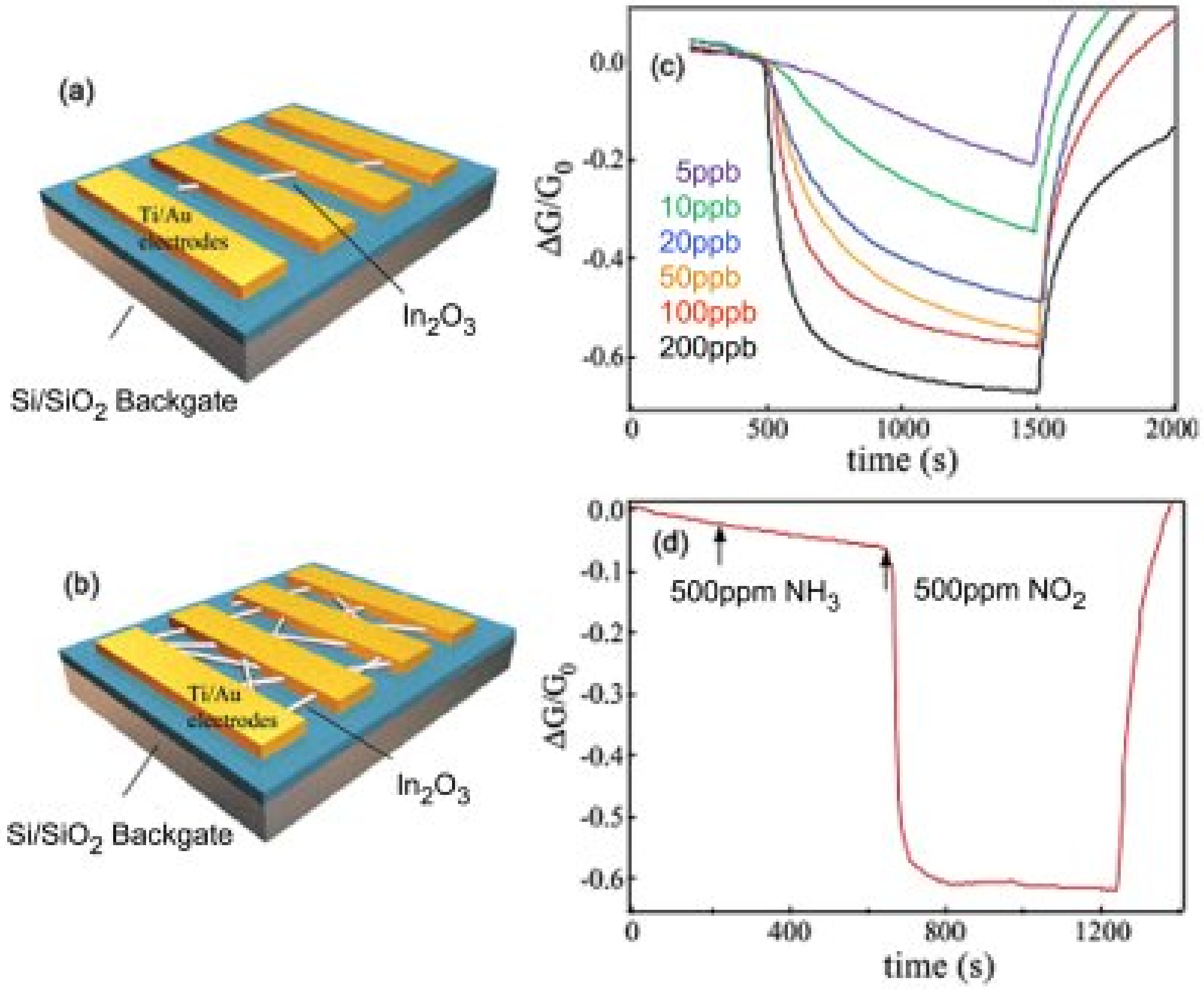

Figure 6. $In_2O_3$ nanowire devices contacted with gold electrodes, (a) single nanowire device (b) multiple nanowire devices, (c) $In_2O_3$ nanowire sensors show a change in conductance when exposed to various concentrations of $NO_2$, and (d) a comparison of response to 500ppm $NH_3$ and 500ppm $NO_2$ (from Ref. [3]).

Multiple nanowire devices also exhibited better selectivity in these experiments. When exposed to $NH_3$ and $NO_2$ at the same concentration, multiwire devices responded strongly to $NO_2$ but weakly to $NH_3$ (Figure 6d). While single $In_2O_3$ nanowire devices showed strongly varying responses that even differed in sign. This was likely caused by the uneven distribution of dopants among different nanowires. When multiple of these single nanowires were contacted as one device, their responses were averaged, which resulted in a very small response to $NH_3$. However the similar phenomenon wasn't observed for $NO_2$. All the single nanowire sensors showed similar and strong response to $NO_2$.

Other types of metal oxide nanowires also showed very impressive performances. It was reported, for example, chemiresistive $V_2O_5$ nanowires/fibers [5] could detect some analytes in a level of ppb. These nanowires/fibers were synthesized by self-assembly in aqueous solution [5]. They were several micrometers long and ~10nm in diameter (see Figure 7A). As sensors, $V_2O_5$ nanowires/fibers could operate at room temperature since they were highly conductive at room temperature. When exposed to analytes, they selectively responded to different analytes. Extremely high sensitivity was measured for 1-butylamine (with a low detection limit of 30ppb, Figure 7 B) and moderate sensitivity for ammonia. In contrast, only very little sensitivity was observed for toluene and 1- propanol vapors [5].

### 3.2 Conducting Polymer Nanowire Sensors

Conducting polymers (CP) are sensitive to gaseous organic vapors at room temperature. Compared to metal oxide sensors, CP sensors thus consume a lower power and require simpler electronic setups. But conducting polymer thin film sensors usually can only sense gaseous analytes at the level of 10 ppm [69,70], which is insufficient for many sensing applications. Polymer nanowires/nanofibers have a higher surface-to-volume ratio which is expected to enhance their sensitivity to vapors. Polymer nanofibers formed by electrodeposition [67,71] and by electrospinning [72] are reported to show enhanced sensitivity to vapor analytes compared to thin films of the same material. Sensitivity to $NH_3$ can be as low as 500ppb [73] with response time on the order of seconds.

Polymer sensors frequently are more chemically sensitive than sensors constructed from metals, semiconductors, or metal oxides. Because of this, contact fabrication cannot be done using standard lithography technologies since these often involve polymer resist and UV or electron beams that will likely damage the chemical structure of the polymer. To avoid lithography, one option is to prefabricate the contact electrodes and then spread the nanowires/nanofibers over the electrodes [74]. The disadvantage of this approach is that the contact resistance varies strongly from material to material and can be large, although there are exceptions, e.g. Polyaniline [74]. In our own laboratory, PEDOT/PSS nanowire devices contacted in this way had very large contact resistance, rendering them unsuitable for use as vapor sensors.

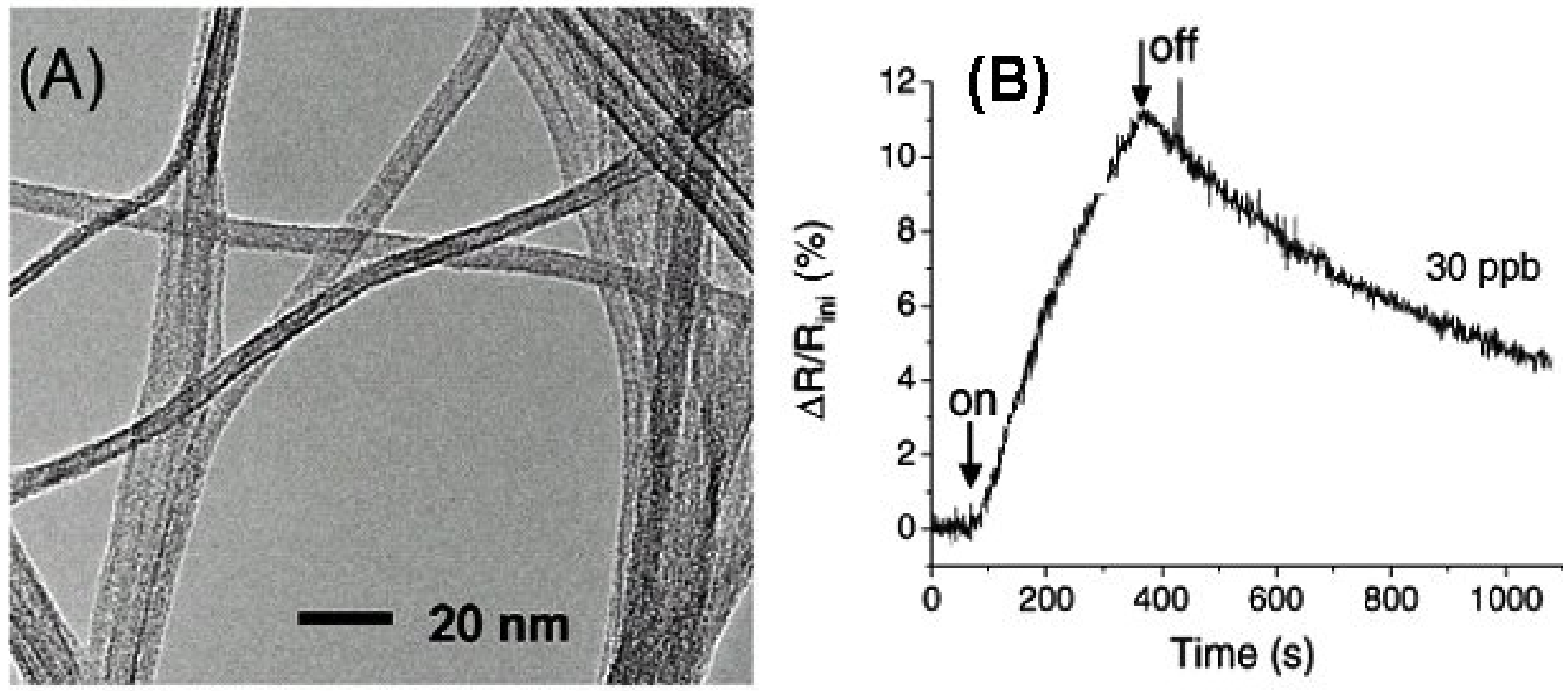

Figure 7. (A) SEM image of $V_2O_5$ nanowires/fibers; (B) Response to 30ppb 1-butylamine (from Ref. [5]).

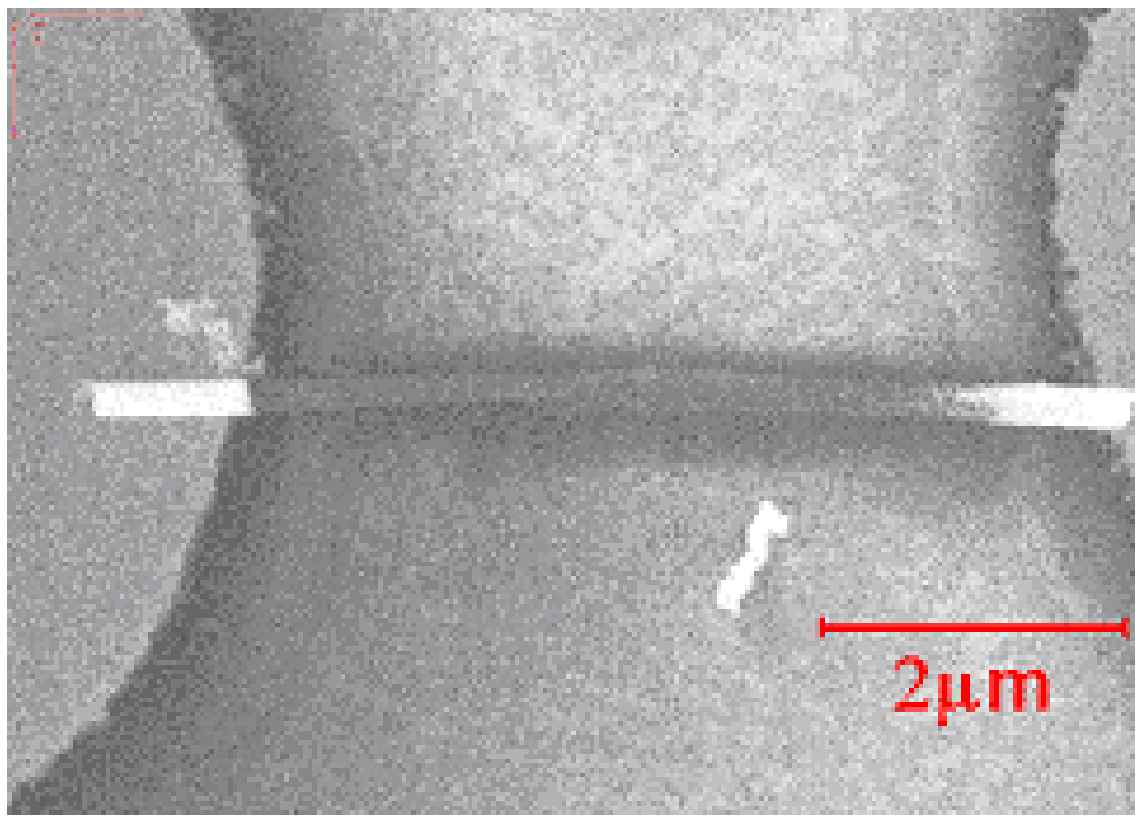

Figure 8. A striped nanowire (gold-polymer-gold) is assembled onto a pair of gold electrodes by dielectrophoresis.

One way to avoid chemical or radiation damage to the polymer nanowire is to use a patterned shadow mask and thermal evaporation so that lithography is avoided entirely. Critical dimensions well below 1 □m are achievable [72,75-78]. Care is needed to ensure that heat generated from evaporating may does not lead to degraded sensing performance of the polymer nanodevices. More importantly, fabrication of large arrays of nanowires can be problematic using either prefabricated electrodes or shadow masks because it is not trivial to ensure proper alignment of the contacts and the nanowires.

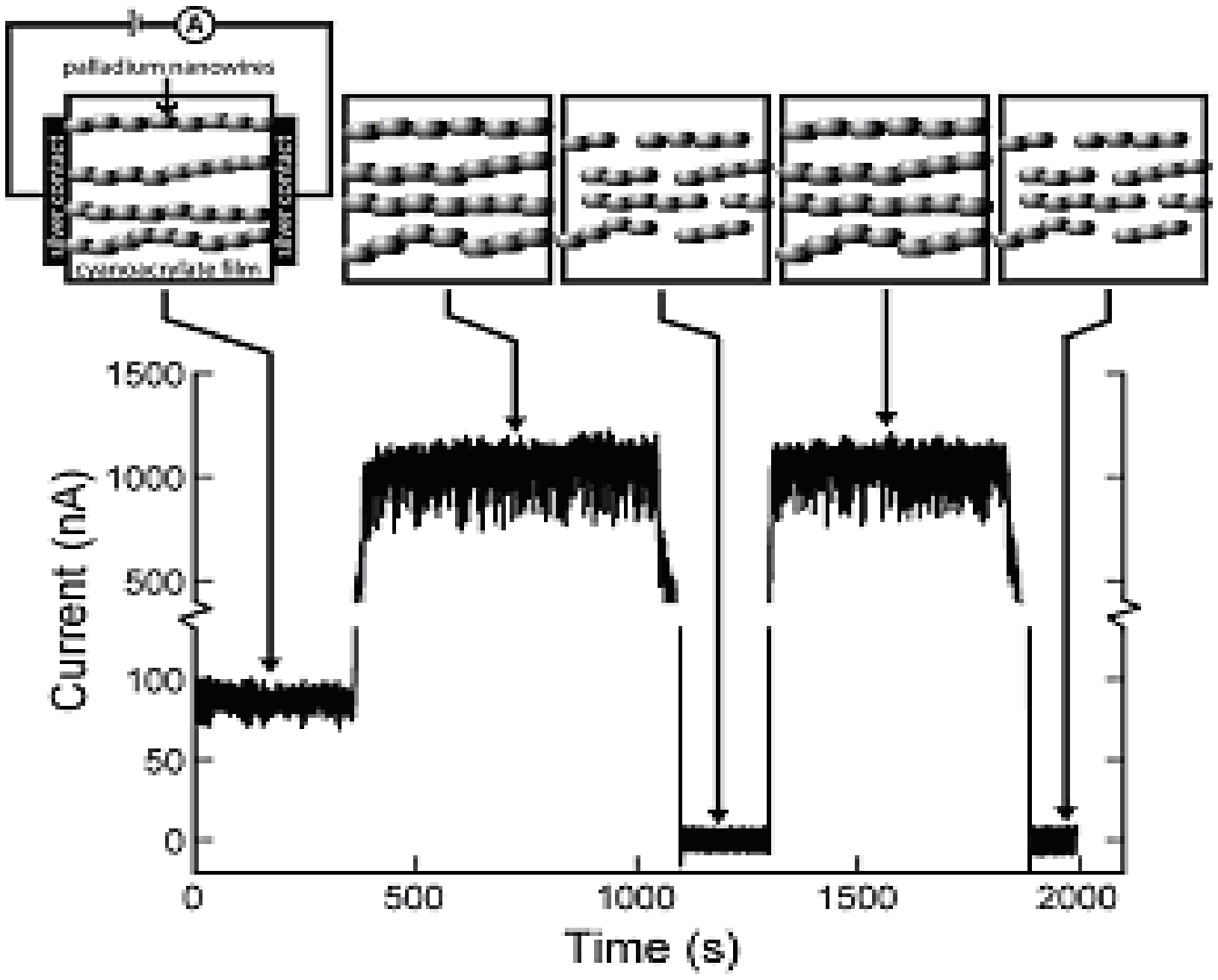

Figure 9. Pd nanowires/mesowires was compressed along the wire axis, which results in a current increase (from Ref. [60]).

In order to solve these problems, nanowires with a "striped" structure (gold-polymer-gold) were grown using a nanoporous template and multiple electrodeposition steps. The striped nanowires were then dielectrophoretically assembled onto prefabricated gold electrodes [9] (Figure 8). The contact resistance between gold and polymer was intrinsic and relatively small since they were electrochemically synthesized. The two gold ends of the nanowire resulted in excellent, reproducible contact with prefabricated gold electrodes. This method guaranteed a minimum contact resistance and was successfully used to make an array of devices consisting of single nanowires.

### 3.3 Metal Nanowire Sensors

Metal Nanowires were found to be sensitive gaseous analytes only in recent years [60-66]. In the year 2000, it was [60] first reported Pd meso/nanowires changed their conductivity upon exposure to $H_2$, due to a structure change of the nanowire itself. These meso/nanowires deposited by electroplating were rough and granular. The dimensions of the grains in the meso/nanowires ranged from 10nm to 300nm. When exposed to $H_2$, the grains in polycrystalline Pd nano/mesowires switched their lattice phase from α to β preferentially at grain boundaries. This compressed the wires along its axis and resulted in the lowering of the intergranular resistance and an increased conductance for each wire, shown Figure 9. The process was repeatable and happened on a very rapid timescale of 20ms. Similar behaviors have been observed for meso/nanowires of Ag and Au [63-66].

### 3.4 Si Nanowire Sensors

It will be quite interesting if silicon nanowires (SiNW) are found to be high quality vapor sensors, given the sophisticated state of silicon fabrication technology. More important, nanowire sensors based on silicon would be readily integrated with sophisticated CMOS integrated circuits that could be used for signal processing and analysis. SiNWs can exhibit excellent field effect transistor characteristics, [14,32] making it likely that SiNW sensors will have high sensitivity. In a recent advance, highly ordered arrays of silicon nanowires (~18nm in diameter) were fabricated using super lattice nanowire pattern transfer technique [4] (Figure 10a). The carrier mobility was as high as $100cm^2V^{-1}s^{-1}$ comparable to that of bulk silicon field transistors. Due to this property, the carrier depletion or accumulation within the nanowires upon exposure to analytes will significantly change the channel conductance. The experimental data showed the nanowire arrays could detect $NO_2$ down to a level of 20ppb, shown in Figure 10b.

The surface of the nanowire sensors is always critical to their performance and surface modification by chemical functionalization has been shown to significantly enhance sensitivity [57]. As recently as 2001, it was demonstrated that antigen-functionalized silicon nanowires could detect molecules down to several picomolar [43].

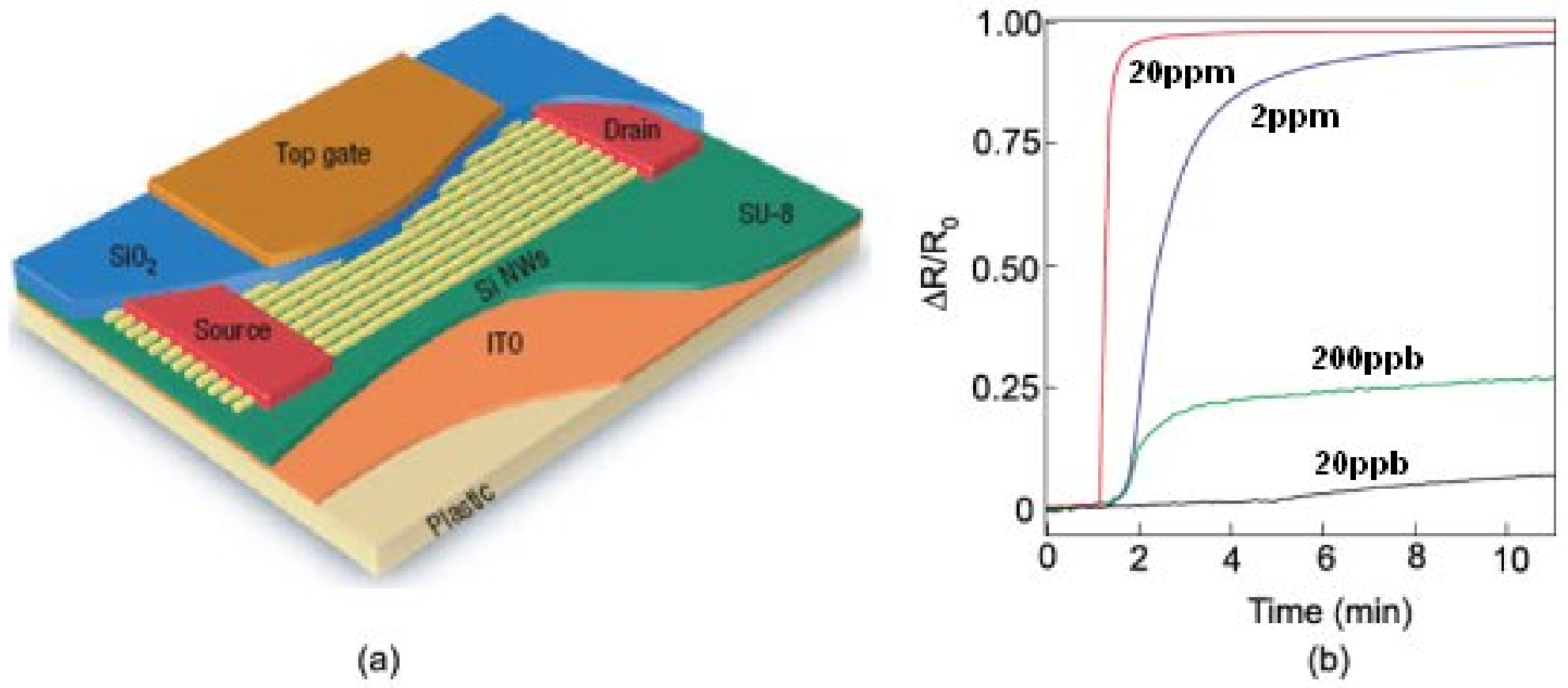

Figure 10. An array of silicon nanowire transistors is shown (a) and their response to $NO_2$ in (b) (from Ref. [4]).

In summary, we introduced the performance of four types of nanowire sensors here: metal oxide, conducting polymers, metals and silicon. Other materials have been used to fabricate nanowire sensors, including semiconducting CdS [49,50].

## 4. Chemiresistors, Transistor Sensors and their Sensing Mechanism

In addition to chemiresistors and transistor sensors, there exist many other types of gas sensors based on different mechanisms. For example, molecule adsorption on the sensor surface will change the surface states which will alter their optical properties. Analytes can be detected by monitoring the optical spectra [79]. Another kind of interesting sensors detects analytes based on resonating frequencies. When molecules are adsorbed onto the surface of a sensor that is incorporated on a resonator, the mass of the sensor will be slightly changed, which shifts the resonant frequency of the resonator. Since resonators are widely used in the modern Micro/Nano electronic mechanical systems, it will be easy to integrate and commercialize this type of chemical gas sensors [80]. Finally nanosized devices are easily made a non-ohmic contact with electrodes [68]. Non-Ohmic contacts are involved in electron transport which is affected by the molecular adsorption. J. R. Morante et al [68] demonstrated an interesting sensor based on the contact resistance.

All these sensing modalities are good options but are not the focus of the current review.

### 4.1 Chemiresistors

When a sensor is exposed to analytes, molecules of the analytes will be chemically adsorbed on the sensor surface. This adsorption alters the surface states and hence changes the sensor resistance. Sensors that detect analytes by monitoring this change are called chemiresistors. Obviously, increasing the surface-to-volume ratio or enhancing the surface adsorption will increase the resistance change and improve the sensitivity.

Nanowires are such excellent nanosized devices for sensing applications due to their high surface-to-volume ratio, simple synthesis process, ease of forming electrical contacts, and quasi one-dimensional carrier confinement. However, compared to thin film chemiresistors, nanowires are found to have comparable a sensitivity until their diameter is smaller than around 25nm [68]. In fact, most of these thin film sensors are wideband metal oxides in the form of polycrystalline with nanosized grains. In this case, the interface between the grains forms a potential barrier which electrons must overcome to effect transport from one grain to another. Adsorption of analyte molecules on the grain boundaries alters the surface state and this potential barrier, which changes the conductance. Therefore smaller grains (limited by specific crystal structures and synthesis conditions) and hence larger surface-to-volume ratio result in a higher sensitivity. For nanowires, only when their diameter is comparable to or smaller than the grain size of the thin film sensor can we observe a higher sensitivity. Electrons inside nanowires of this small size have a quasi one-dimensional transport which also contributes to this higher sensitivity [81].

Another approach to increase the sensitivity is to enhance the molecule adsorption by modifying the surface. For example, $SnO_2$ nanowires doped with Pd particles exhibit significantly improved sensitivity [57] due to the formation of Schottky barrier junctions induced by Pd particles on the nanowire surface. The adsorption and desorption of analyte molecules such as $O_2$ modulate this Schottky junction, which alters the overall resistance of the nanowire. At the same time, Pd particles also act as catalysts that can pre-dissociate the adsorbed molecules to be atomic species. These atomic species are much easier to interact with the surface, which therefore increases the sensitivity of the sensors. More interestingly, instead of doping with catalyst particles, the sensitivity of nanowire sensors can be improved by functionalizing with bio-molecules, as discussed later.

In either case, the response of chemiresistors is found to follow a power law, especially for metal oxide and conducting polymer sensors:

$$\frac{\Delta R}{R_0} = \alpha P^{\beta}$$

where P is the partial pressure of analytes, and $\alpha$ and $\beta$ are constants.

For metal oxide sensors, $\beta$ typically has a rational fraction value, usually 1 or ½, depending on the charge of the surface species and the stoichiometry of the elementary reactions on the surface [47]. $\beta$ is also affected to a certain degree by other factors such as nanowire diameter and working temperature. In Figure 11, $\beta$ for $SnO_2$ nanowires (~60nm in diameter) is found to increase from 0.48-0.58 for CO exposure as the working temperature is increased from 200°C to 280°C. This behavior recently has been explained theoretically by combining the depletion theory of semiconductor, which deals with the dynamics of electronics between surface state and bulk, with the dynamics of adsorption and/or reactions of gases on the surface [82].

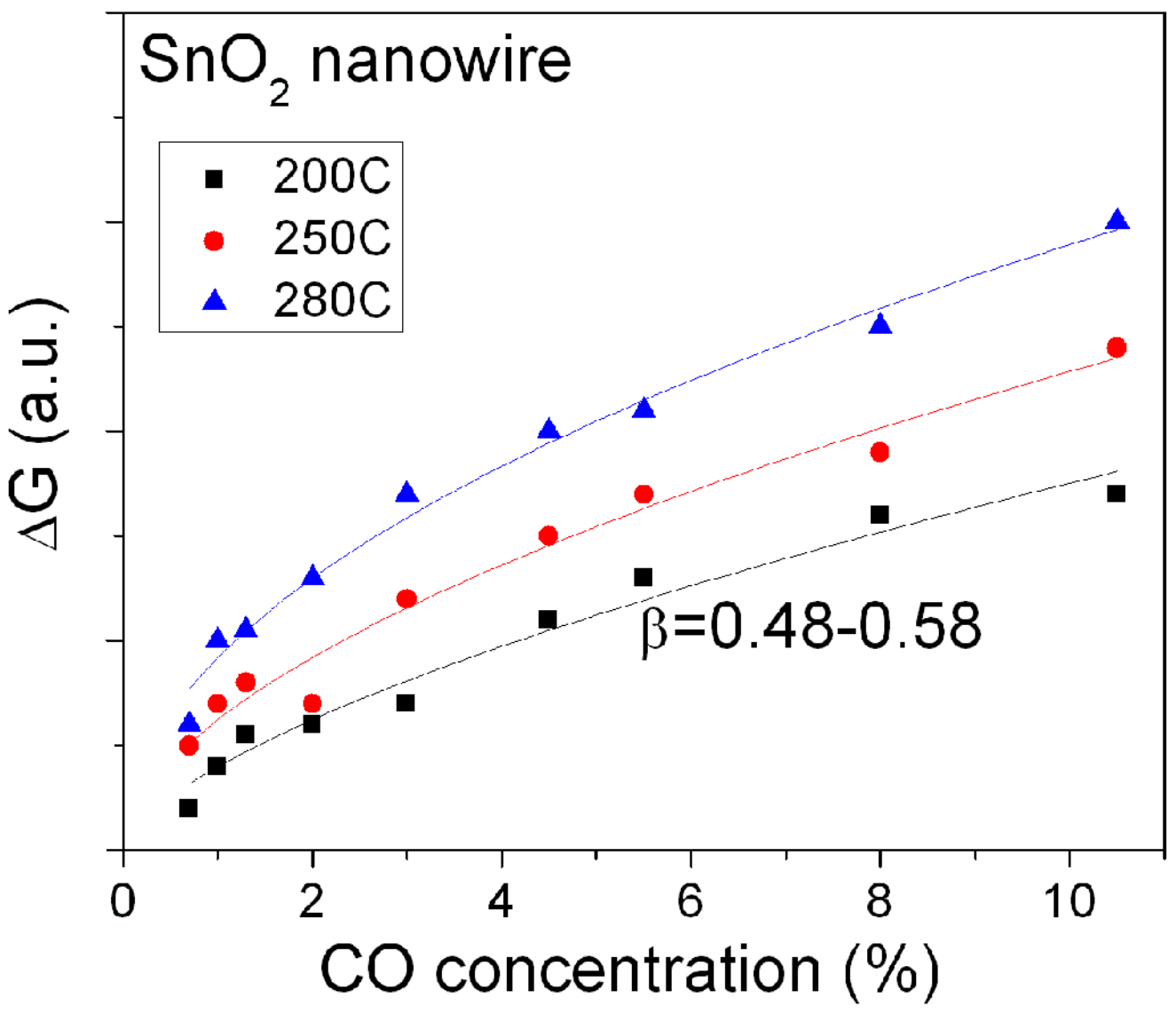

Figure 11. The conductance change of SnO2 nanowires follows a power law of CO concentration (from Ref. [47]).

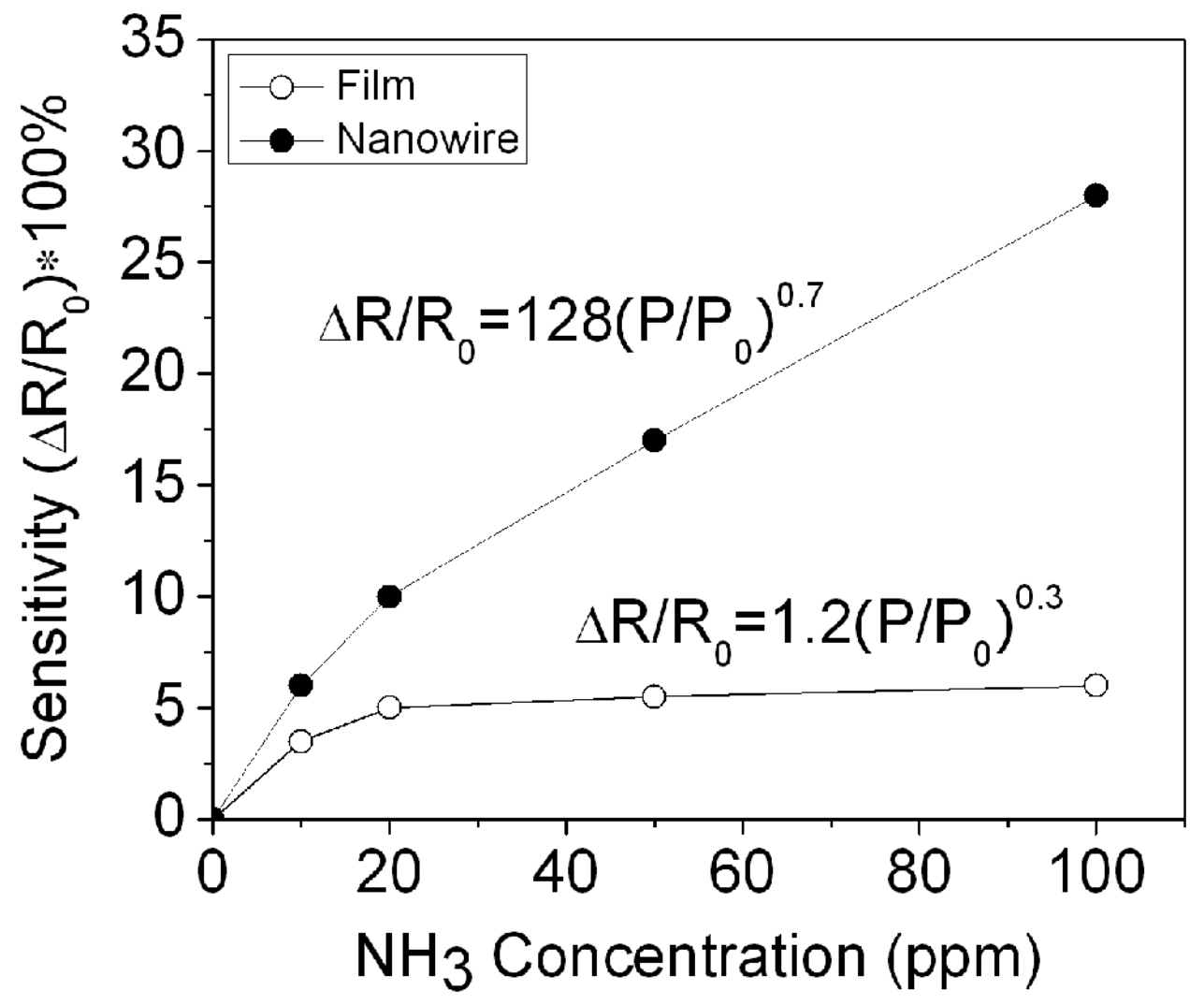

Figure 12. Sensitivities of Poly (3,4-ethylenedioxythiophene) nanorods and films upon exposure to $NH_3$ (from Ref. [83]).

We have observed similar power laws for conducting polymer sensors. Unlike metal oxide sensors, β for conducting polymer sensors can vary over a wide range, depending on the composition and diameter of the nanowire and the working temperature. Our experimental data [84] showed the β value for PEDOT/PSS nanowires increased from 0.8 to 1.4 as the working temperature was elevated from room temperature to 60°C. Also, polymer nanowire sensors usually have a higher β value than films of the same material, as shown in Figure 12. However, the sensing mechanism for conducting polymer sensors is still in dispute. Some [85] claimed polymers swelled after exposed to analytes, which moved apart the polymer chain from each other and the conductance of the polymer therefore decreased [85]. Others claimed charge transfer was involved in this process. [83]

### 4.2 Transistor Sensors

Transistors have long been used as sensors to detect gaseous analytes. The gate effect of the transistors can amplify the modulation of the channel conduction. Molecule adsorption on the surface that involves charge transfer acts like a gate on the transistors, which is called the chemical gating effect [86] (Figure 13a). Nanowires are perfect for this kind of sensing applications. They can be synthesized as a single crystal (thin films are typically polycrystalline with grain boundaries) that has a uniform and continuous channel. For example, modifying the silicon oxide surface of a single crystal silicon nanowire (SiNW) with 3-amino-propyltriethoxysilane (APTES) can chemically gate the SiNW, shown in Figure 13a. APTES molecules are covalently linked to SiNW oxide surface which results in a surface terminating in both $–NH_2$ and –SiOH groups. At low pH, the $–NH_2$ group is protonated to $–NH_3^+$ and acts as a positive gate, which depletes hole carriers in the p-type SiNW and decreases the conductance. At high pH, -SiOH deprotonated to $–SiO^-$ correspondingly causes an increase in conductance [86]. It is known that the gating effect can amplify the channel conductance change. Thus it is expected that such nanowire transistors have a high sensitivity. Indeed, biotin functionalized silicon nanowires can detect streptavidin down to several picomolar [86]. Also, single crystal $In_2O_3$ nanowire transistors are also be able to detect $NO_2$ in a level of several ppb [2,3].

More interestingly, when a gate voltage is applied on the nanowire transistors (for instance, through a back gate), the gate voltage will significantly enhance or depress the molecule adsorption on the transistor surface [11,24,46]. In Figure 13b, a $SnO_2$ nanowire transistor is used to sense $O_2$ and CO (at a constant concentration) while the back gate voltage is changing from 0 to –6 volts. During this change, the n-type channel conductance in $N_2$ is constantly decreasing. A more negative gate voltage depresses the ion-adsorption of $O_2$, which diminishes the $O_2$ induced decrease in conductance. The ion adsorption of CO on the surface of $SnO_2$ is more complicated. It's believed that at least two interaction processes are involved here, such as CO with $O^{2-}$ in $SnO_2$, and CO with surface hydroxyl groups ($Sn^+OH^-$). These complicated interactions induce an initial increase and a later decrease in conductance change when the gate voltage moves from 0 to more negative values.

## 5. ASSEMBLY TECHNOLOGIES

The scale of nanowire arrays is critical for the application of electronic noses. Numerous efforts have been taken to develop methods to integrate nanowires in a large array at low cost [87,84,88]. Here we discuss some of the more promising recent advances in this area.

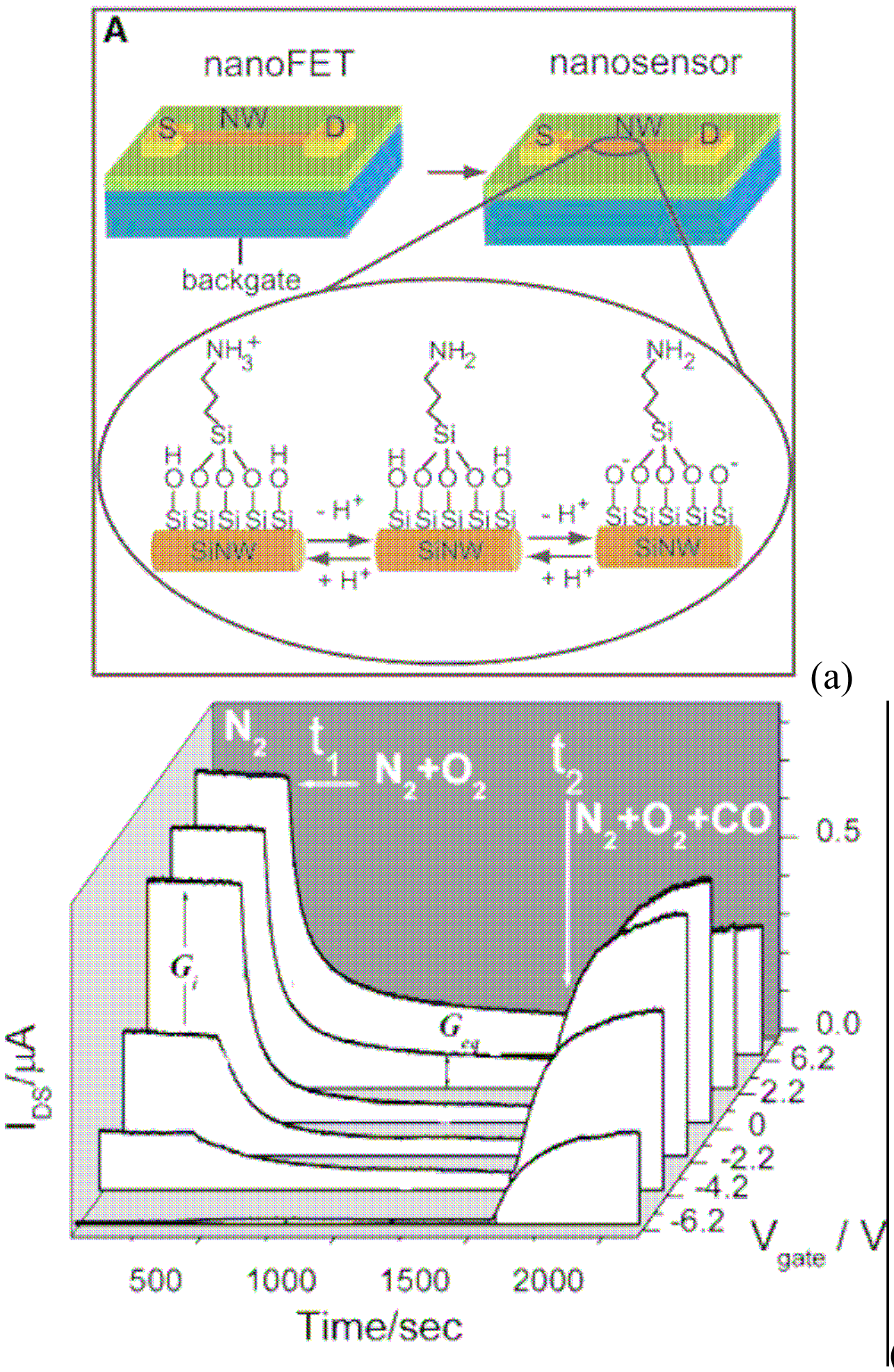

(a)

(b)

Figure 13. (a) Changes in the surface charge state of the APTES-modified SiNW surface with pH (from Ref. [86]); (b) The channel conductance changes with the gate voltage upon exposure to $O_2$ followed by CO (from Ref. [24]).

## 5.1 Self-assembly

Nanoporous membrane templates using diblock copolymer composed of polystyrene (PS) and polymethylmethacrylate (PMMA) was first reported [87] in the year 2000. To fabricate such templates, the diblock copolymer film was first self-assembled into cylindrical microdomains under the guidance of external electric fields when the temperature was kept above the glass transition temperatures of both PS and PMMA. Then the PMMA domain was degraded and the PS matrix cross-linked upon exposure to deep ultraviolet beams. In the end, the PMMA domain was removed with solvents and thus left a nanoporous structure in the membrane. This membrane, with a high density of periodically distributed nanopores, has features that can compete with the commercial nanoporous anodic aluminum oxide membranes (AAO).

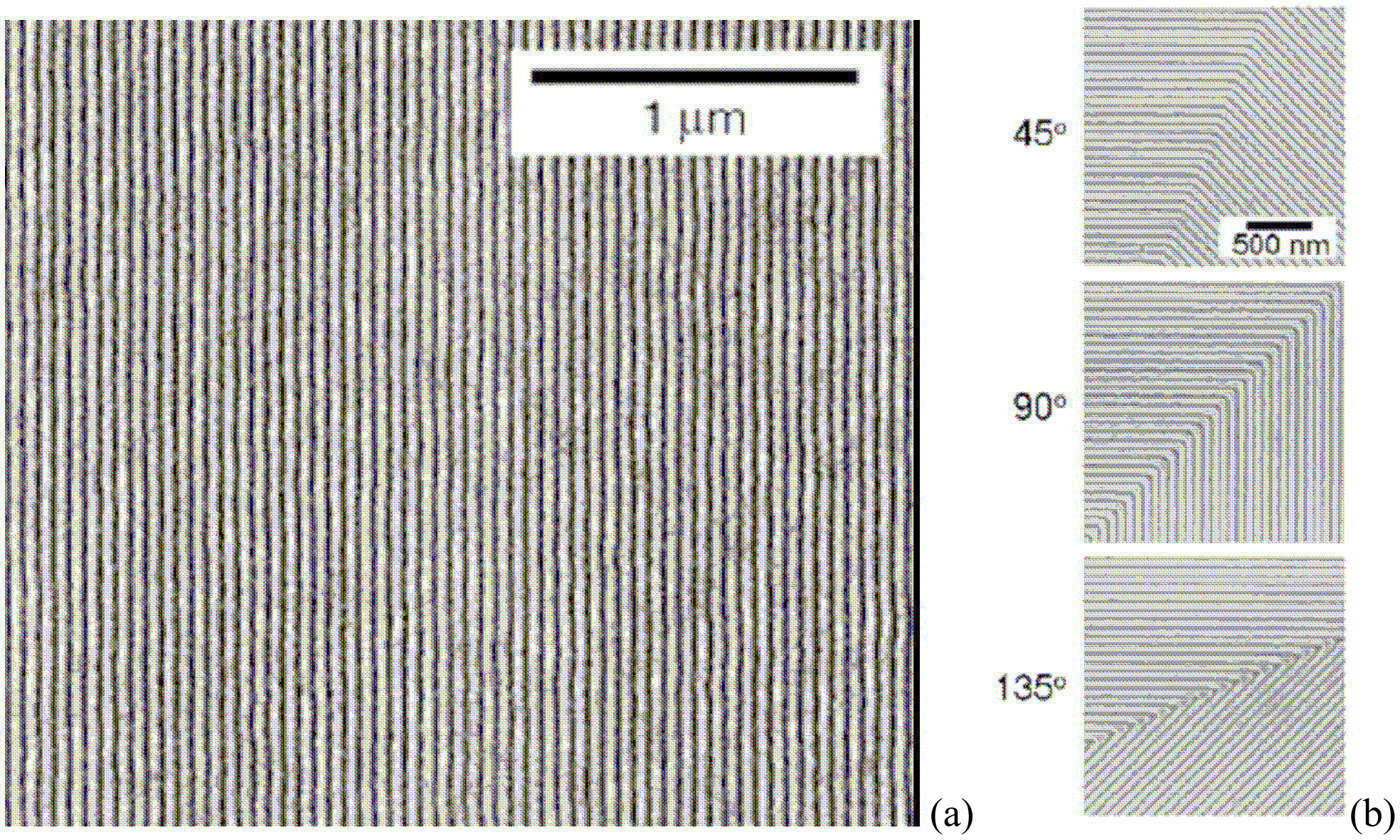

Figure 14. Copolymer arrays with straight line and angled lines. The bright and dark regions correspond to the PS and PMMA ((a) from Ref. [6] and (b) from Ref. [7]).

More interestingly, an inexpensive and scaleable technique for patterning dense periodic arrays of parallel lines was developed using the same diblock copolymer (Figure 14a). Later nonregular device-oriented structures were created using this technique [7], shown in Figure 14b where the white and black lines represent the polystyrene and PMMA, respectively.

To date these approaches have not been used for the fabrication of nanowire sensors, but this is certainly an intriguing possibility. In addition to the fabrication of nanowires, self-assembly has already been used to create 3-D structures [89] more recently.

### 5.2 Lithographic Patterning

Today's advance lithography technologies such as deep UV photolithography, E-beam lithography and Nanoimprint lithography have been able to make nano-size devices. Due to their controllability and flexibility, it is quite reasonable to use these technologies to make large arrays of nanoscale sensors. It was reported last year that Nanoimprint lithography was employed [90] to create dense arrays of silicon nanowires over a large area, shown in Figure 15a. When exposed to analytes such as ammonia gas, the threshold voltage of the field-effect transistor was shifted. More recently, a titania polycrystalline nanowire array for gas sensing applications was reported [91] (Figure 15b). A solid thin film of titania was first deposited using the sol-gel technique and the film was then patterned into an array of nanowires by the side etching effect of SF6 plasma after photolithography. The nanowire array was found to be much more sensitive to ethyl alcohol vapor than the thin film.

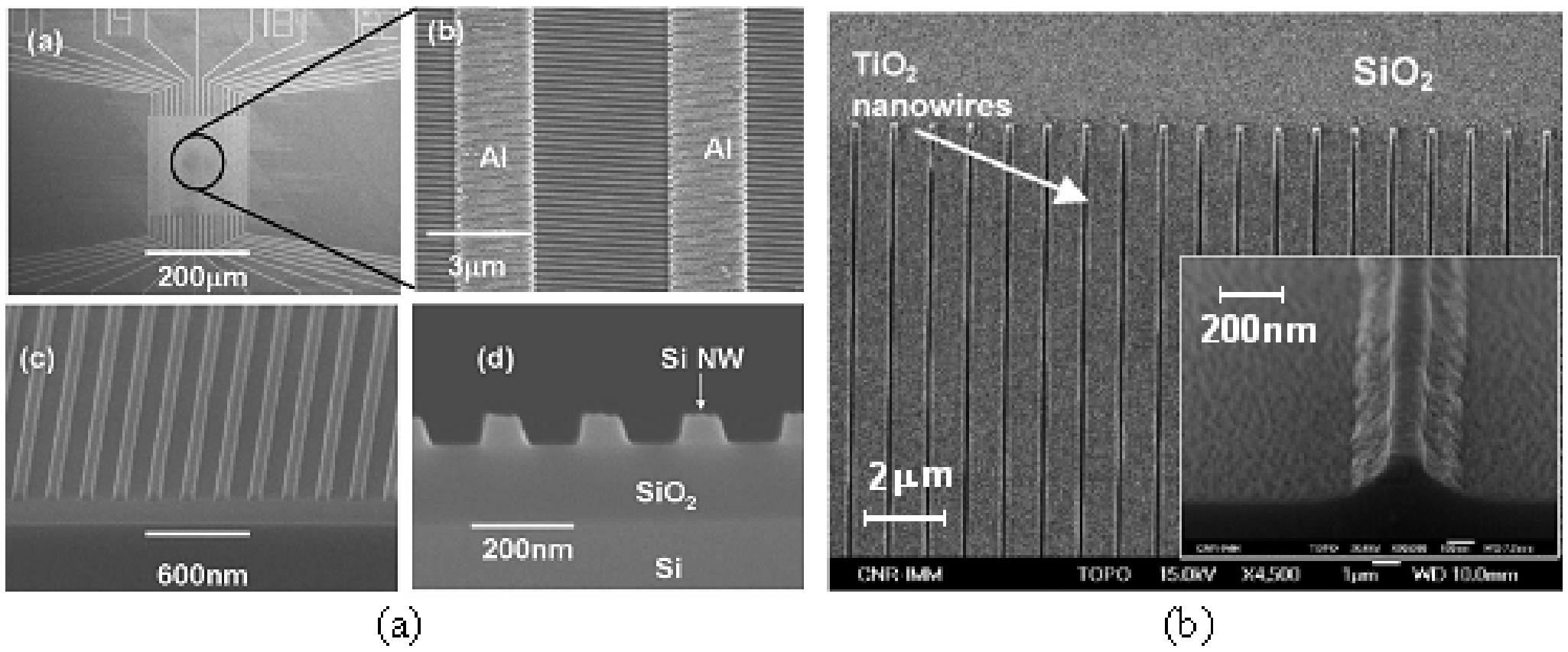

Figure 15. (a) Silicon nanowire arrays made from photolithography (from Ref. [90]) and (b) $TiO_2$ nanowire arrays made from a sputtered $TiO_2$ thin film (from Ref. [91]).

### 5.3 Dielectrophoretic Assembly

Dielectrophoretic assembly is an excellent method to assemble nanosized devices from a solution suspension onto specific locations, guided by an electric field gradient produced by an applied AC voltage. This assembly method relies on the fact that a polarizable particle (metal, semiconductor, or insulator) experiences a net force in a DC electric field gradient, even if its net charge is identically zero. However, in the experiments considered here, the particle is suspended in a solution (for example, water with a particular ion concentration) that is also polarized in this DC field. The polarized solution offsets the polarization of the particle, which means the force on the particle will be reduced or in some cases even reversed in sign. However, if we use an ac voltage source with such a high frequency that the solution media cannot be polarized in pace with the ac voltage source, the media then will not be polarized any more. The force on the polarized particle resumes its original value.

To characterize the properties of a single nanowire device, we need to assemble one nanowire per site. A *self-limiting* assembly process is thus desirable [9,92]. One simple method of achieving this goal is described below.

In the experiments, the device assembly site (Figure 16) consists of a pair of electrodes separated by approximately 5 μm (the assembly capacitor). One of the two electrodes is capacitively coupled by another 5 μm gap (the coupling capacitor) to a feed-in lead, and the second electrode is grounded. The coupling capacitor and assembly capacitor are designed to be approximately 200 fF and 10 fF, respectively (Figure 16a). A voltage of 20V peak-to-peak at a frequency of 100 kHz is applied across these two capacitors whose impedances at this frequency (8MΩ and 160MΩ, respectively) are 100–1000 times greater than that of a conducting nanowire (supposed to be ~100 kΩ). These impedance values imply that before assembly about 19V drops across the assembly capacitor, leading to a high electric field in this gap that draws nanowires into the desired location. After one nanowire is brought into contact with both electrodes, the voltage drop across the assembly capacitor decreases to 200mV, so the assembly process is self-terminating. With samples consisting of an array of 30 such electrode pairs, an assembly rate of 50% is obtainable. After assembly, a small drop of silver paste is used to short the coupling capacitor and enable dc electrical measurements.

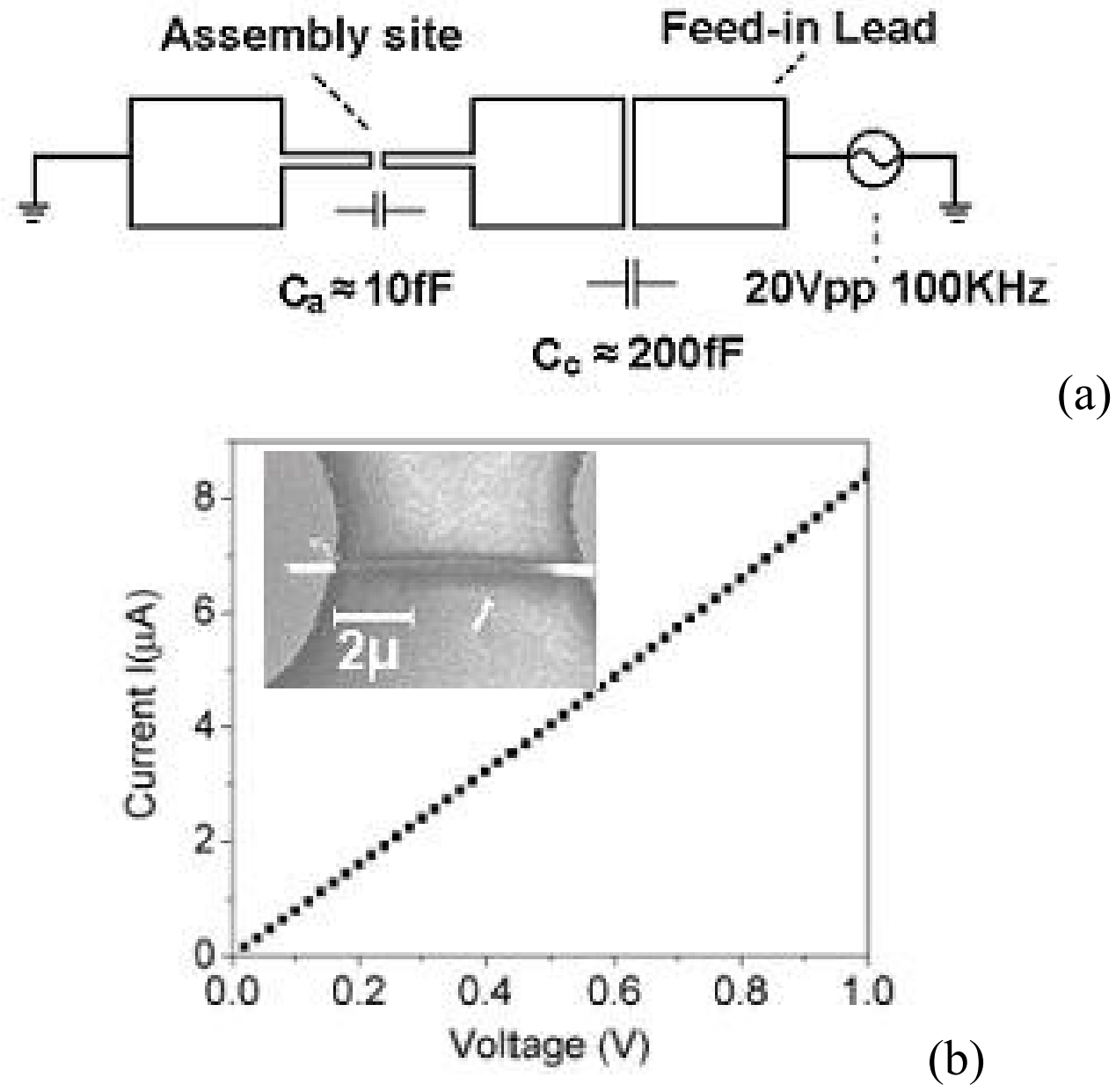

Figure 16. A) Schematic of circuitry used for self-limiting dielectrophoretic assembly. B) Representative current- voltage curve from a single nanowire device. Inset: A striped polymer nanowire bridging a pair of gold electrodes (white ends are gold and between them is the polymer part).

## 5.4 Water Assembly and Blown Bubble Assembly

In recent years, quite a few new approaches building nanowire arrays have been proposed and demonstrated [93,94]. One of the approaches uses the Langmuir-Blodgett (LB) technique to uniaxially compress the nanowire-surfactant monolayer on an aqueous subphase (Figure 17). This method can align nanowires with a controllable spacing that can be later transferred to the entire surface of substrates.

The spacing of the well-aligned nanowires has a scale of micrometers to submicrometers (~200nm). Compression of spacing below 200nm leads to increasing aggregation due to strong nanowire-nanowire attraction. Aggregated Core-Shell nanowires with the shell as the sacrifice layer can be used to effect the assembly of nanowire arrays with much lower spacing (~50nm, for example) after the shell sacrifice layer is removed [93].

Another interesting approach to create large arrays of aligned nanowires, called Blown Bubble Film method, was recently reported [94]. With this method, nanowires were first introduced into an organic solvent (such as 5,6-epoxyl-triethoxysilane) to form a viscous suspension. At a controlled pressure, a bubble was formed out of the suspension under the guidance of a circular die which controlled the vertical expansion rate. The result was a large number of nanowires aligned along one direction (±10$^{o}$) on the exterior surface of the bubble. These nanowires could then be transferred onto a substrate for further processing.

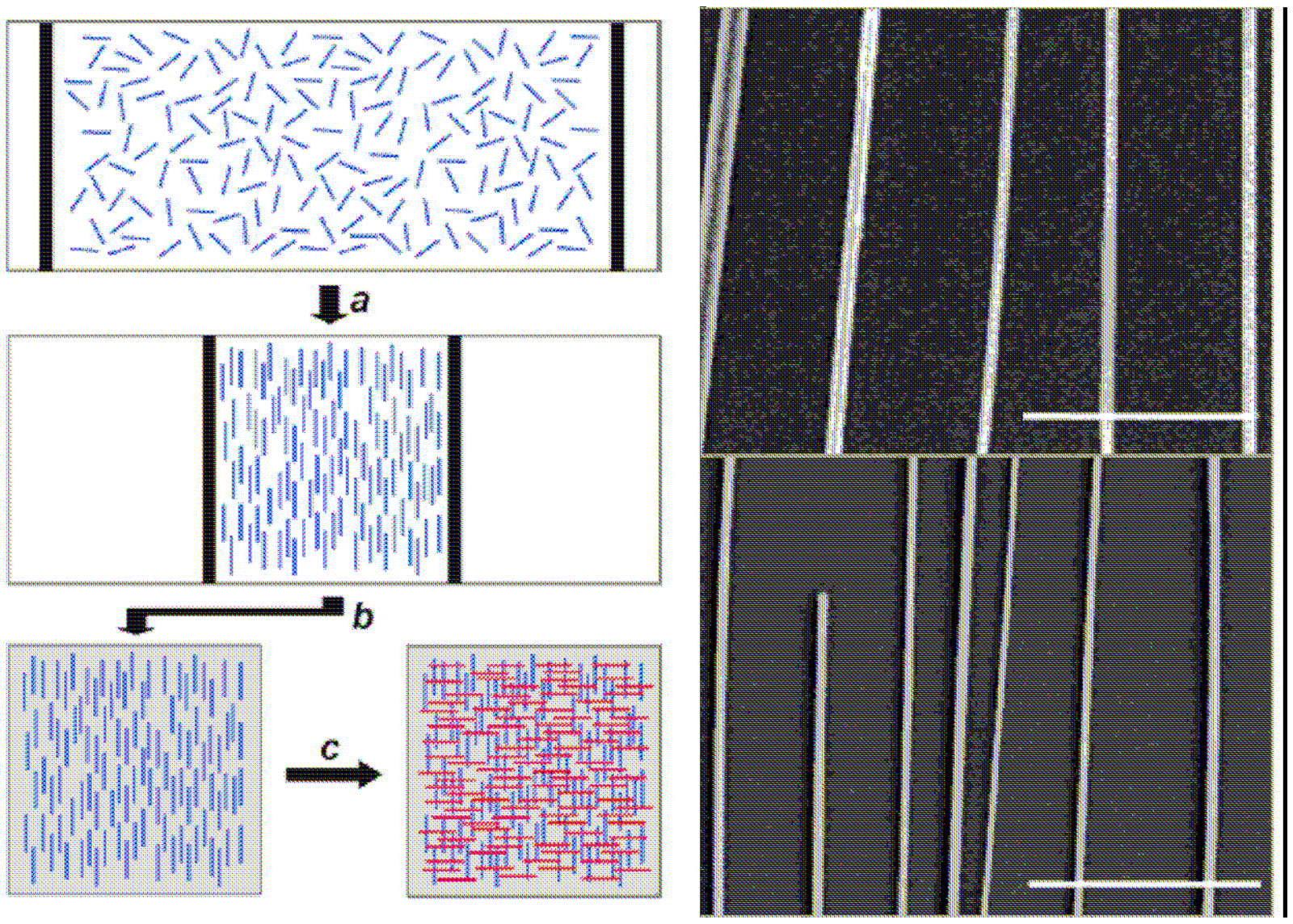

Figure 17. Nanowires (blue lines) in a monolayer of surfactant at the air-water interface are (a) compressed on a Langmuir-Blodgett trough to a specified pitch. (b) The aligned nanowires are transferred to the surface of a substrate to make a uniform parallel array. (c) Crossed nanowire structures are formed by uniform transfer of a second layer of aligned parallel nanowires (red lines) perpendicular to the first layer (blue lines). The left figure is a SEM image of aligned nanowires (from Ref. [93]).

In addition to the methods we discussed, there are many other approaches to create nanowire arrays, such as optical trapping and patterned catalyst assisted LVS, etc. Optical trapping, the so-called optical tweezers, has a long history and its principle is well understood [95]. It is a powerful way to manipulate single nano devices. Although as commonly implemented this method is a serial assembly process and not applicable for assembling a large scale of nano devices, parallel approaches based on "holographic" optical tweezers have been developed [97]. From the synthesis section in this chapter, it's widely known catalysts will control the growth of nanowires. The distribution of catalysts patterned in a specific way may offer an interesting approach to fabricate nanowire arrays in a large scale. This idea was recently demonstrated [88] by grow well-oriented and size-controlled Si nanowire arrays.

## 6. Future Perspective

An extremely large-scale array with enough diversity and high sensitivity is the fundamental requirement for "electronic noses". Research on sensors has been on this theme for decades. Before the age of micro-technology, sensors can be built only in a macro scale. A large-scale array with diversity is almost impossible. More important, the sensitivity of macro-sensors is typically low. Research then was focused on understanding the sensing mechanism and developing new process, new materials and new dopants to increase the sensitivity. In the age of micro-technology, with improved sensitivity, researchers started developing sensor arrays and various pattern recognition techniques to process data, in addition to research on process, materials and dopants.

In the era of nanotechnology, we are about to be ready for building "electronic noses", considering tremendous progress on chemical sensors have been achieved. 10 nm nanowire sensors have been able to detect analytes in a level of several ppb. A lot of new materials are found to good sensors such as Pd, Au, Ag and conducting polymers. More important, various self-assembly techniques for nanowire sensor arrays have been discovered, such as diblock-copolymer self-assembly, water assembly and blown bubble assembly. Using these assembly techniques, we are ready to build large sensor arrays with different types of nanowires, which will be a hot topic next decades. To build "electronic noses", people still need to find out more efficient assembly techniques, more sensitive sensors and even new sensing materials. In terms of sensitivity, bio-functionalization is a very interesting and promising approach. As for new sensor devices, a nanoparticle sandwiched in a nanowire operating as a short channel nano-transistor will be an interesting new type of sensors. A sensor with this structure may have faster response and higher sensitivity compared to similar nanowire sensors.